\documentclass{emulateapj}
\usepackage{graphicx}
\usepackage[space]{grffile}
\usepackage{latexsym}
\usepackage{textcomp}
\usepackage{longtable}
\usepackage{multirow,booktabs}
\usepackage{amsfonts,amsmath,amssymb}
\usepackage{natbib, verbatim}
\usepackage{url}
\usepackage{hyperref}
\hypersetup{colorlinks=false,pdfborder={0 0 0}}
\newif\iflatexml\latexmlfalse
\usepackage[utf8]{inputenc}
\usepackage[english]{babel}
\usepackage{todonotes}

\usepackage{tikz}

\newcommand\pc{{\, \rm pc}}

\newcommand\myr{{\, \rm Myr}}

\newcommand{\kms}{\,\rm{km}\,\rm{s}^{-1}} 
\newcommand\msun{{\, \rm M_\odot}}

\usepackage{xspace}

\newcommand\GN{GN\xspace}
\newcommand\GNs{GNs\xspace}
\newcommand\GC{GC\xspace}
\newcommand\SMBH{SMBH\xspace}
\newcommand\SMBHs{SMBHs\xspace}
\newcommand\NSC{NSC\xspace}
\newcommand\NSCs{NSCs\xspace}
\newcommand\CW{CW\xspace}
\newcommand\CNR{CNR\xspace}
\newcommand\SPH{SPH\xspace}

\newcommand{\myfloatalign}{\centering}

\begin{document}


\title{ Forming circumnuclear disks and rings in galactic nuclei: \\a competition between supermassive black hole and nuclear star cluster}
\shorttitle{Circumnuclear gas morphology as indicator of SMBH presence}


\author{Alessandro A. Trani\altaffilmark{1,2,3,$\star$,$\dagger$}}

\author{Michela Mapelli\altaffilmark{3,4}}

\author{Alessandro Ballone\altaffilmark{3}}

\shortauthors{Trani et al.}

\altaffiltext{1}{Department of Astronomy, Graduate School of Science, The University of Tokyo, 7-3-1 Hongo, Bunkyo-ku, Tokyo, 113-0033, Japan}
\altaffiltext{2}{Scuola Internazionale Superiore di Studi Avanzati (SISSA), Via Bonomea 265, I--34136, Trieste, Italy}
\altaffiltext{3}{INAF-Osservatorio Astronomico di Padova, Vicolo dell'Osservatorio 5, I--35122, Padova, Italy}
\altaffiltext{4}{Institute for Astrophysics and Particle Physics, University of Innsbruck, Technikerstrasse 25/8, A–6020, Innsbruck, Austria}

\altaffiltext{$\star$}{Email: aatrani@gmail.com}
\altaffiltext{$\dagger$}{JSPS Fellow}


\begin{abstract}



We investigate the formation of circumnuclear gas structures from the tidal disruption of molecular clouds in galactic nuclei, by means of smoothed particle hydrodynamics simulations.
We model galactic nuclei as composed of a supermassive black hole (SMBH) and a nuclear star cluster (NSC) and consider different mass ratios between the two components.
We find that the relative masses of the SMBH and the NSC have a deep impact on the morphology of the circumnuclear gas.
Extended disks form only inside the sphere of influence of the SMBH. In contrast, compact rings naturally form outside the SMBH's sphere of influence, where the gravity is dominated by the NSC. This result is in agreement with the properties of the Milky Way’s circumnuclear ring, which orbits outside the SMBH sphere of influence. 
{
Our results indicate that compact circumnuclear rings can naturally form outside the SMBH sphere of influence. 
}

\end{abstract}

\bibliographystyle{apj}



\keywords{ISM: clouds -- black hole physics -- ISM: kinematics and dynamics -- methods: numerical -- galaxies: nuclei -- galaxies: star clusters: general}


\section{Introduction}

Galactic nuclei (GNs) can be remarkably rich in molecular gas. 
An increasing number of observations reveal the presence of circumnuclear gas in the innermost parsecs of nearby \GNs, where the gravity is dominated by a supermassive black hole (SMBH, \citealt{neu07,tri09,set10,dav13,men13,men15,oni15,bar16a,bar16b,oni17,dav17,esp17}). 

Molecular gas is also present in Milky Way's Galactic center (GC), where a dusty molecular torus - the so-called circumnuclear ring\footnote{\footnotesize In the literature the \CNR can be sometimes referred as "circumnuclear disk" or "torus". In our manuscript, we define a flattened gas structure as a "ring" if it has a significant hole in its center. This definition was adopted to simplify the notation in the manuscript, i.e. to clearly distinguish between a disk (without hole) and a ring. According to e.g. \citet{liu13} the \CNR extends from ${\sim}1.5$--$2\pc$ to $3$--$4\pc$ and it has a clear "hole" in the centre. In addition, the \CNR has an asymmetric, 7pc-long extension towards negative galactic longitude. This extension may be a remnant of the parent molecular cloud that has been captured and disrupted by the potential well at the nucleus. For example, see figure 7 of Takekawa et al. 2017.
}
 (CNR) - orbits at about $2\pc$ from the \SMBH \citep{ser86,wri01,chr05,oka11,liu12,liu13,mil13,goi13,smi14,har15,tak17,mil17,san17,goi18}.

Circumnuclear gas exhibits a complex morphology and kinematics, with clumpy streamers, warped rings and/or disks that deviate from axisymmetry and circular motion. How the complex spatial and velocity structure of circumnuclear gas forms and evolves remains poorly understood. 
This uncertainty limits the use of molecular gas dynamics to infer the dynamical mass of \SMBHs, a method that is recently emerging thanks to high-resolution sub-millimeter interferometry \citep{dav14,yoo17}.

Furthermore, circumnuclear gas may form stars in the close proximity to \SMBHs \citep{pol77,kol80,shl87,san98,col99,gam01,lev03,goo03,mil04,nay05,nay07,cua08,col08,yus12,yus13,yus15,yus17,yus18}. Evidence comes from the observations of young stars orbiting around the \SMBH in the GC and in M31 \citep[e.g.][]{ghe03,sch03,eis05,pau06,gil09a,bar09,do13,yel14,fel15,tre95,ben05,lau12,men13,bro13,loc17}. 

Similar episodes of star formation in the central parsecs are expected to be detected in other nearby galaxies with the advent of 30-meter telescopes \citep{gul14,do14,car15}. Young, luminous stars such as the ones we observe in the Milky Way's \GC are excellent candidates to infer the \SMBH mass from kinematic measurements \citep{sch02,ghe03,gil09b}.
Such stars may even retain the dynamical properties of the parent gas and thus help to constrain the past history of the \GN \citep{map17}.

Circumnuclear gas may form from the tidal disruption of giant molecular clouds. This scenario has been studied in detail in the context of the \GC and can explain (i) the formation of the \CNR and (ii) the young stellar disk \citep{war08,bon08,hob09,luc13,tra16,map16a}.
However, little attention has been paid to this scenario in \GNs with different properties from those of the Milky Way.

In general, the outcome of a cloud tidal disruption depends on details of the central gravitational field. 
\SMBHs are commonly thought to dominate the gravitational field of the central parsecs. 
On the other hand, many GNs host a massive and compact stellar cluster at their center - the so-called nuclear star cluster (\NSC) -, whose mass is also a fundamental component of the central potential well. NSCs and SMBHs may even co-exist in the same GN.

In fact, many \GNs host a compact, massive stellar cluster at their center, and both \NSCs and \SMBHs may coexist in the same \GN \citep{bok02,bok04,cot06,geo14,bro14,geo16,ngu18}.

The most striking example is in the \GC, where the \NSC contains twice the mass of the \SMBH at only ${\sim}2\pc$ \citep{sch07,gra09,sch09,yus12,sch14,fel14,cha15,fri16,gal17,sch17,fel17}. 

The mass ratio between \SMBH and \NSC can vary widely from galaxy to galaxy, consequently affecting the shape of the central potential. This is expected to have a strong impact in the disruption of nearby molecular clouds and hence might shape the properties of circumnuclear gas.


In this paper, we present the first systematic study on the formation of gaseous circumnuclear rings/disks in \GNs with properties different from those of the \GC.
We simulate the infall of a molecular cloud towards the central potential of \GNs, composed of a \SMBH and the cusp of a \NSC. 
We run a grid of smoothed-particle hydrodynamics (SPH) simulations by varying the mass ratio between the \SMBH and the stellar cusp, and study the properties of the resulting distribution of gas.

Section~\ref{sec:methods} describes the numerical methods we employed for our simulations. Section \ref{sec:results} presents the main results from the simulations. In Section~\ref{sec:discussion}, we discuss the impact of the \SMBH to \NSC mass ratios on circumnuclear gas morphology and its implications for \SMBH mass measurement. Finally, our conclusions are summarized in Section~\ref{sec:conclusions}.

\section{Methods}\label{sec:methods}

We use the N-body/\SPH code {\textsc gasoline2} \citep{wad04,rea10,wad17} to simulate the infall and disruption of a molecular cloud in the central parsecs of \GNs. 

We consider \GNs as composed of a stellar cusp and a central \SMBH. The \SMBH is a sink particle of mass $M_{\rm SMBH}$ whose position is fixed at the center; in this way we avoid spurious random walk due to numerical effects.

The stellar cusp of the \NSC is modeled as a spherical potential and follows a broken power-law density profile:
\begin{equation}\label{eq:cusp}
\rho(r) = \rho_0 \left( \frac{r}{r_0} \right)^{-\gamma}
\end{equation}
We use the values given by \citet{sch07} for the cusp of the Milky Way: $\gamma = 1.75$ for $r>r_0$ and $\gamma = 1.2$ for $r < r_0$, where $r_0 = 0.22\pc$. The cusp is truncated at $r_{\rm trunc}=100\pc$ and has a total mass $M_{\rm cusp}$.
We choose $\rho_0$ in the following way: first we choose the \SMBH mass ($M_{\rm SMBH}$) and the total mass of the \GN within $10\pc$ [$M_{\rm tot} = M_{\rm cusp}({<}10\pc) + M_{\rm SMBH}$]. Then we pick $\rho_0$ so that $M_{\rm cusp}({<}10\pc) = \int_0^{10\pc} 4\pi\rho(r)r^2\rm{d}r = M_{\rm tot} - M_{\rm SMBH}$.

We have run three sets of simulations, each with a different value of $M_{\rm tot}$: $5\times10^7$, $1\times10^7$ and $5\times10^6 \msun$. 
For each set we have run four simulations choosing $M_{\rm SMBH}$ so that $f_{\rm SMBH} = M_{\rm SMBH} / M_{\rm tot} = 0.5, 0.2, 0.1$ and $0.05$. This choice allows us explore the parameter space of \SMBH and \NSC masses in a range consistent with the observations \citep{set08,gra09,kor09,kor13,geo14,geo16}.

\makeatletter\onecolumngrid@push\makeatother
\begin{table*}

  \caption{Main properties of the simulations.\label{tab:ic}}
  \begin{minipage}{\linewidth}
    \begin{center}
    \begin{tabular}{lcrlccc}
      
	Run & $M_{\rm tot}$ [$\msun$] & $M_{\rm SMBH}$ [$\msun$] & $f_{\rm SMBH}$ & $R_{\rm SOI}$ [$\pc$] & $m_{\rm res}$ [$\msun$] 	\vspace{3pt} \\\hline

	\texttt{mt5e7\_bh2.5e7} & $5\times10^7$ & $2.5\times10^7$ & $0.5$ & $>$$10$ & $0.5$\\ 
	\texttt{mt5e7\_bh1e7} & $5\times10^7$ & $1\times10^7$ & $0.2$ & $0.90$ & $0.5$\\
	\texttt{mt5e7\_bh5e6} & $5\times10^7$ & $5\times10^6$ & $0.1$ & $0.35$ & $0.5$\\\vspace{2pt}
	\texttt{mt5e7\_bh2.5e6} & $5\times10^7$ & $2.5\times10^6$ & $0.05$ & $0.15$ & $0.5$\\	
	
	\texttt{mt1e7\_bh5e6} & $1\times10^7$ & $5\times10^6$ & $0.5$ & $>$$10$ & $0.5$\\ 
	\texttt{mt1e7\_bh2e6} & $1\times10^7$ & $2\times10^6$ & $0.2$ & $0.90$ & $0.5$\\
	\texttt{mt1e7\_bh1e6} & $1\times10^7$ & $1\times10^6$ & $0.1$ & $0.35$ & $0.5$\\\vspace{2pt}
	\texttt{mt1e7\_bh5e5} & $1\times10^7$ & $5\times10^5$ & $0.05$ & $0.15$ & $0.5$\\

	\texttt{mt5e6\_bh2.5e6} & $5\times10^6$ & $2.5\times10^6$ & $0.5$ & $>$$10$ & $0.5$\\
	\texttt{mt5e6\_bh1e6} & $5\times10^6$ & $1\times10^6$ & $0.2$ & $0.90$ & $0.5$\\
	\texttt{mt5e6\_bh5e5} & $5\times10^6$ & $5\times10^5$ & $0.1$ & $0.35$ & $0.5$\\\vspace{2pt}
	\texttt{mt5e6\_bh2.5e5} & $5\times10^6$ & $2.5\times10^5$ & $0.05$ & $0.15$ & $0.5$\\

	\texttt{mt1e7\_bh5e6\_hr} & $1\times10^7$ & $5\times10^6$ & $0.5$ & $>$$10$ & $0.05$\\ 
	\texttt{mt5e6\_bh1e6\_hr} & $5\times10^6$ & $1\times10^6$ & $0.2$ & $0.35$ & $0.05$ \\
	\texttt{mt5e7\_bh1e6\_hr} & $5\times10^7$ & $1\times10^6$ & $0.02$ & $0.06$ & $0.05$ \\
      \end{tabular}
 \end{center}\vspace{3pt}
    \end{minipage}
	{\footnotesize 
	  Column~1: run name; column~2: mass enclosed in a $10\pc$ radius $M_{\rm tot}$ in $\msun$, composed of the NSC and the \SMBH; column~3: mass of the \SMBH $M_{\rm SMBH}$ in $\msun$; column~4: $f_{\rm SMBH}$; column~5: radius of sphere of influence $R_{\rm SOI}$ of the \SMBH in $\pc$; $m_{\rm res}$ is the mass resolution of the simulation in $\msun$.}
\end{table*}
\makeatletter\onecolumngrid@pop\makeatother

From the values of the \SMBH mass and the \NSC cusp profile, we derive the radius of the sphere of influence of the \SMBH $R_{\rm SOI}$. Although usually the sphere of influence is defined as the region enclosing a total mass twice that of the \SMBH mass, this definition is valid only for an isothermal sphere model of the \NSC.
Therefore, $R_{\rm SOI}$ is computed numerically from the equation $\Phi_{\rm SMBH} =  \Phi_{\rm cusp}$, where $\Phi_{\rm SMBH}$ and $\Phi_{\rm cusp}$ are the gravitational potential of the \SMBH and of the cusp, respectively.
Table~\ref{tab:ic} summarizes the main properties of the simulations presented in this paper.

In all simulations, the molecular cloud is modeled as a homogeneous gas sphere of $10^5\msun$ and $15\pc$ radius, located at $26\pc$ from the \SMBH.
The cloud is seeded with turbulent velocity and marginally self-bound. The velocity field is generated using a grid method \citep{dub95} from a divergence-free, random Gaussian field with a power spectrum $P(k) = \left\| \delta v_k \right\|^2 \propto k^{-4}$. The spectral index $-4$ reproduces turbulence in agreement with the velocity dispersion relation $\sigma_{v} \propto l^{0.5}$ observed in molecular clouds \citep{lar81}.

The cloud has an impact parameter of $b=15\pc$ with respect to the \SMBH and an initial velocity of $v_{\rm i}=0.2 v_{\rm esc}$, where $v_{\rm esc}$ is the escape velocity of the cloud, taking into account the potential of the \SMBH and the \NSC. The initial velocity is $v_{\rm i} = 46.7$, $20.1$ and $13.7 \kms$
in the runs with $M_{\rm tot} = 5\times10^7$, $1\times10^7$ and $5\times10^6 \msun$, respectively. 
The coordinate system is chosen so that the total velocity vector lies in the $x$-$y$ plane. 
The mass of the gas particles is $m_{\rm res} = 0.05\msun$ in high-resolution runs and $0.5\msun$ in every other runs. We stop the simulations at $3 \myr$.


Star formation is modeled via sink particle creation, implemented following the criteria of \citet{bat95} and \citet{fed10}.
Specifically, a gas particle is considered a sink candidate if it exceeds a threshold density $\rho_{\rm thr}$.
We choose $\rho_{\rm thr} = 10^{-16} \rm\,g\,cm^{-3}$. We performed tests to ensure that different values of $\rho_{\rm thr}$ do not impact significantly on the mass function of the formed sink particles. We set a sink accretion radius of $r_{\rm accr} = 1.5\times10^{-3}\pc$.
The gravitational softening length is chosen to be $7.93 \times 10^{-4} \pc$ and $3.68 \times 10^{-4}\pc$ in runs with $m_{\rm res} = 0.5\msun$ and $m_{\rm res} = 0.05\msun$, respectively.

We employ the \citet{cul10} viscosity limiter, which uses the total time derivative of the velocity divergence as shock indicator. 


All simulations include the radiative cooling algorithm described in \citet{bol09} and \citet{bol10}. The cooling is calculated from $\nabla \dot F = -\left(36\pi \right)^{1/3} s^{-1} \sigma \left( T^4-T^4_{\rm irr} \right) \left(\Delta\tau + 1/\Delta\tau\right)-1$, where $s = (m/\rho)^{1/3}$ and $\Delta\tau = s k \rho$, for the local opacity $k$, particle mass $m$, and density $\rho$. \citet{dal01} opacities are used, with a $1 \,\text{\textmu}m$ maximum grain size. The irradiation temperature is $T_{\rm irr} = 100\,\rm K$ everywhere.

\makeatletter\onecolumngrid@push\makeatother
\begin{figure*}[!htpb]
  \begin{minipage}{\linewidth}
    \myfloatalign
    \includegraphics[width=\linewidth]{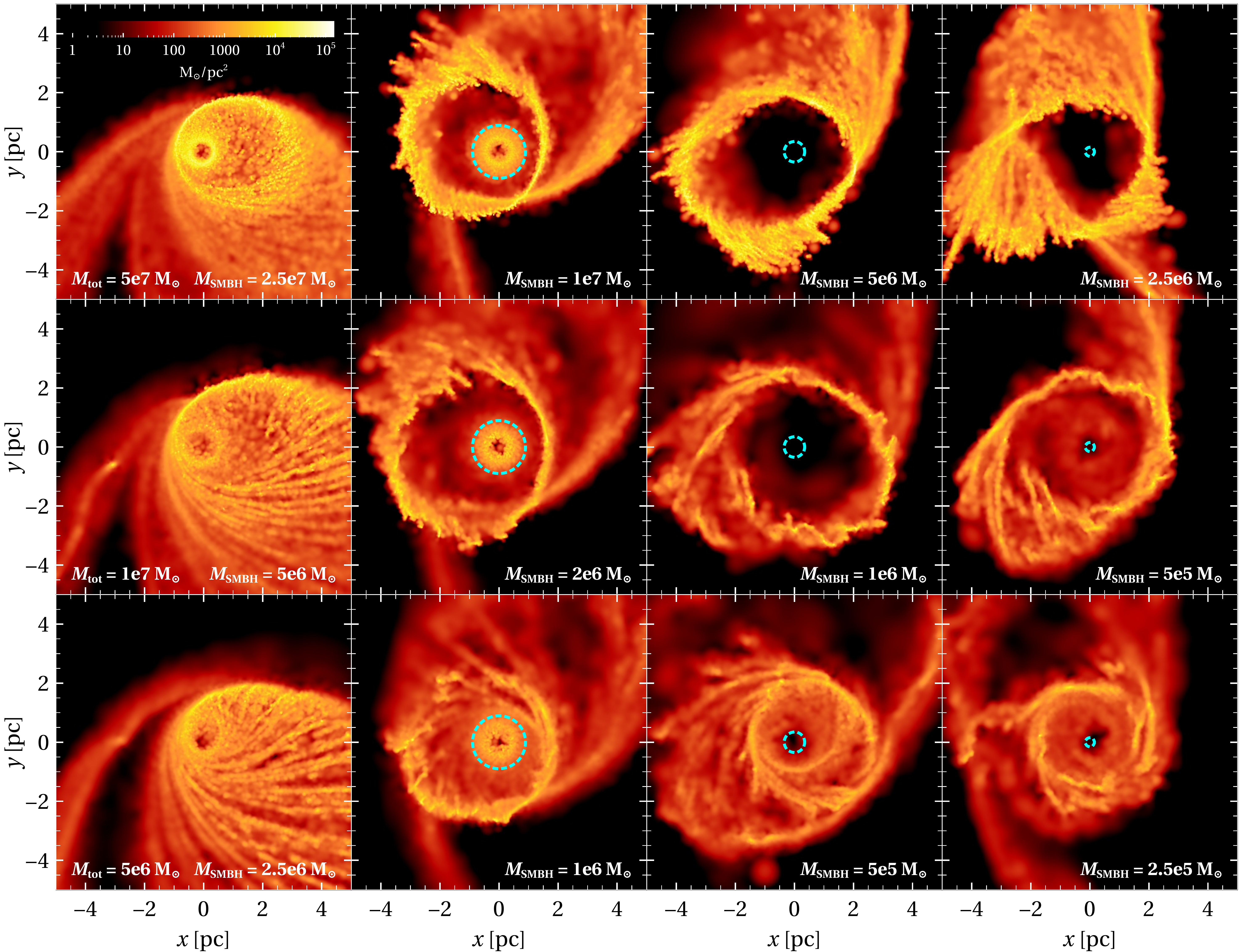}
    \end{minipage}%
	\caption{Color-coded, projected density map of gas in the $x$-$y$ plane for the three sets of runs (rows), comprising of four initial setups (columns, see Table~\ref{tab:ic}). Each panel corresponds to a different simulation. From top to bottom, each row has $M_{\rm tot} = M_{\rm cusp} + M_{\rm SMBH}$ of  $5\times10^7$, $1\times10^7$ and $5\times10^6 \msun$, respectively. From left to right, each column has $f_{\rm SMBH} = M_{\rm SMBH} / M_{\rm tot} = 0.5, 0.2, 0.1$ and $0.05$. The dashed cyan circle indicates the \SMBH radius of influence $R_{\rm SOI}$. From top to bottom, each row corresponds to a time of $0.6$, $1.5$ and $2\myr$ from the start of the simulations. Bottom row snapshots are taken at a later times because the gas cloud evolves more slowly in simulations with lower $M_{\rm tot}$.  $R_{\rm SOI}$ is not shown in the left-hand panels because it is larger than the box. 
	}
	\label{fig:grid}
\end{figure*}
\clearpage
\makeatletter\onecolumngrid@pop\makeatother

\section{Results}\label{sec:results}

{ From our simulations we study the tidal disruption of the cloud in the tidal field of the \SMBH and the \NSC. The contribution of the \SMBH will dominate the potential inside its influence radius. Thus, we expect that gas reaching the influence radius of the \SMBH will settle on a nearly Keplerian orbit around the \SMBH. In contrast, the \NSC will give the dominant contribution to the potential outside the \SMBH influence radius. We expect that gas settling outside the \SMBH influence radius will feel mostly the effect of the \NSC.}

\begin{figure*}[!htpb]
    \begin{minipage}{\linewidth}
	\includegraphics[width=\linewidth]{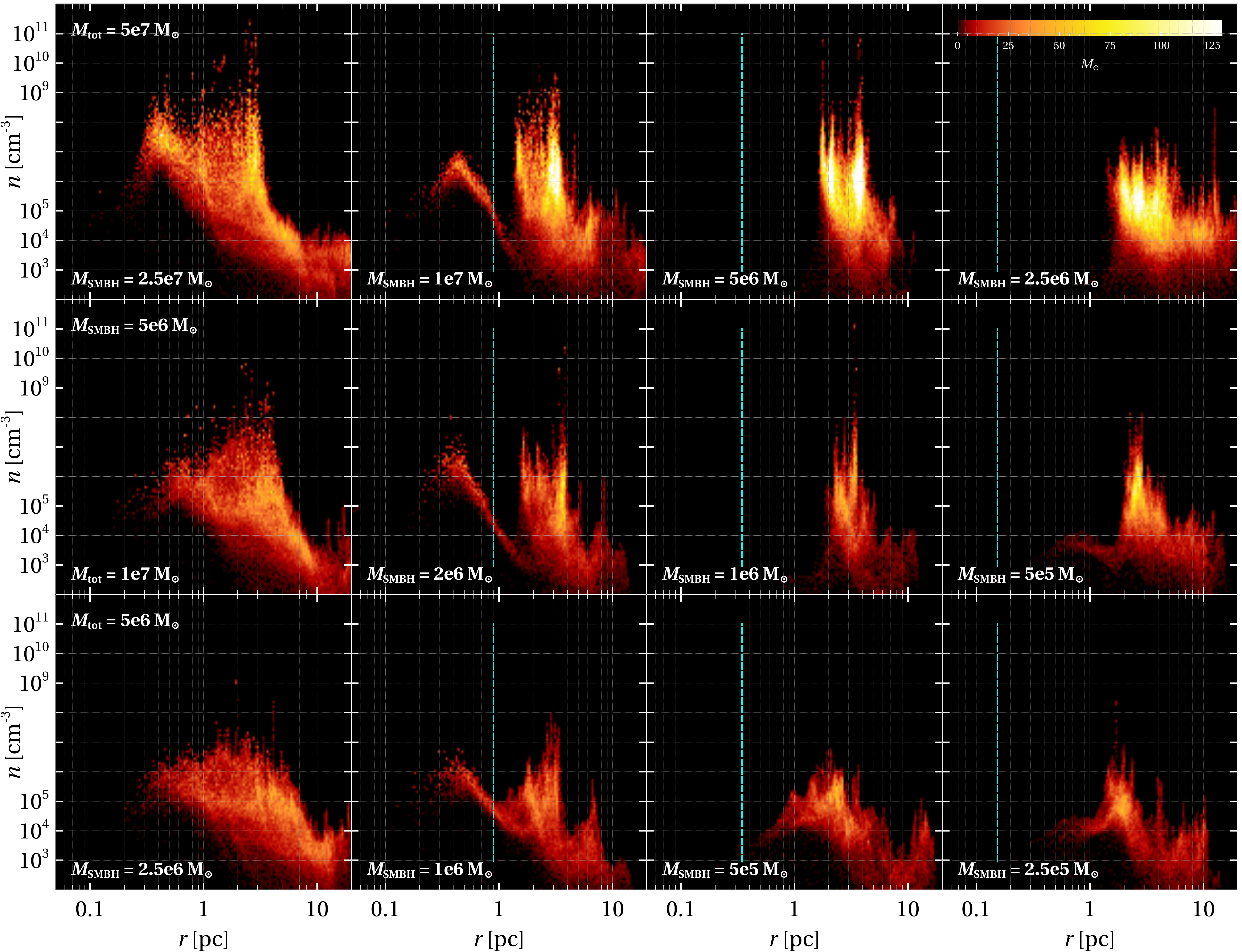}
    \end{minipage}
	\caption{ Number density of gas as a function of distance from the central \SMBH for all snapshots of Figure~\ref{fig:grid}. The color scale indicates the gas mass in the radial-density bin. The cyan dashed line is the \SMBH influence radius $R_{\rm SOI}$. Each panel corresponds to a different simulation. From top to bottom, each row has $M_{\rm tot} = M_{\rm cusp} + M_{\rm SMBH}$ of  $5\times10^7$, $1\times10^7$ and $5\times10^6 \msun$, respectively. From left to right, each column has $f_{\rm SMBH} = M_{\rm SMBH} / M_{\rm tot} = 0.5, 0.2, 0.1$ and $0.05$. From top to bottom, each row corresponds to a time of $0.6$, $1.5$ and $2\myr$ from the start of the simulations. Bottom row snapshots are taken at a later times because the gas cloud evolves more slowly in simulations with lower $M_{\rm tot}$. $R_{\rm SOI}$ is not shown in the left-hand panels because it is larger than the box. 
	}
	\label{fig:omaggio}
\end{figure*}

Figure~\ref{fig:grid} shows the projected density map of the whole grid of simulations at different snapshots. The total mass $M_{\rm tot}$ of the \GN decreases from top to bottom, while the ratio between the \SMBH mass and the total mass of the \GN ($f_{\rm SMBH}$) decreases from left to right. The morphology of the circumnuclear gas shows a clear trend with $f_{\rm SMBH}$. At $f_{\rm SMBH}=0.5$, the whole cloud gets flattened into an eccentric, extended disk around the \SMBH. As the mass of the \SMBH becomes lower with respect to that of the \NSC ($f_{\rm SMBH}\leq0.2$), the gas gets squeezed into a compact ring outside the sphere of influence of the \SMBH. A disk of material captured by the \SMBH potential resides within the cavity of the ring in the runs with $f_{\rm SMBH}=0.2$. This is clear also from Figure~\ref{fig:omaggio}, which shows the corresponding radial distribution of the gas density in for all snapshots of Figure~\ref{fig:grid}.

For $f_{\rm SMBH}<0.2$, the sphere of influence of the \SMBH is too small to allow for the capture of gas particles in our initial condition. A smaller initial velocity and/or impact parameter would lead to the formation of a disk inside $R_{\rm SOI}$ also in these cases \citep{map16a}. 

\makeatletter\onecolumngrid@push\makeatother
\clearpage
\makeatletter\onecolumngrid@pop\makeatother

Table~\ref{tab:gasres} summarizes the properties of the disk and the inner ring formed in the simulations
after the complete disruption of the cloud, which corresponds to $0.6$, $1.5$ and $2 \myr$ from the start of the simulations with  $M_{\rm tot} = 5\times10^7$, $1\times10^7$ and $5\times10^6 \msun$, respectively. 



\makeatletter\onecolumngrid@push\makeatother
\begin{table*}
	
	\caption{Main outcomes of the simulations.\label{tab:gasres}}
	\begin{minipage}{\linewidth}
          \centering
\begin{tabular}{lccccccccc}
        Run &$r^{\rm in}_{\rm disk}$ [pc] & $r^{\rm out}_{\rm disk}$ [pc] & $M_{\rm disk}$ [M$_\odot$] & $r^{\rm in}_{\rm ring}$ [pc] & $r^{\rm out}_{\rm ring}$ [pc] & $M_{\rm ring}$ [M$_\odot$] & $M_{\rm stars}$ [M$_\odot$] & $M_{\rm accr}$ [M$_\odot$] \\ \hline
	
	\texttt{mt5e7\_bh2.5e7} & $0.23$ & $7.5$  & $7.1\times 10^4$ & $-$ & $-$ & $-$ & $1.0\times 10^4$ & $7.0\times 10^3$\\ 
	\texttt{mt5e7\_bh1e7} & $0.2$ & $1.0$ & $4.3\times 10^3$ & $1.4$ & $3.6$ & $4.6\times 10^4$ & $2.9\times 10^4$ & $3.9\times 10^3$\\
	\texttt{mt5e7\_bh5e6} & -- & -- & -- & $1.6$ & $4.0$ & $6.2\times 10^4$ & $2.4\times 10^4$ & $2.2\times 10^3$ \\\vspace{2pt}
	\texttt{mt5e7\_bh2.5e6} & -- & -- & -- &	$1.7$ & $5.6$ & $5.9\times 10^4$ & $2.2\times 10^4$ & $1.7\times 10^3$ \\
	
	\texttt{mt1e7\_bh5e6} & $0.23$ & $9.2$  & $8.5\times 10^4$ & $-$ & $-$ & $-$ & $4.6\times 10^4$ & $6.4\times 10^3$\\ 
	\texttt{mt1e7\_bh2e6} & $0.15$ & $0.9$ & $2.8\times 10^3$ & $1.6$ & $4.0$ & $1.7\times 10^4$ & $7.1 \times 10^4$ & $3.6 \times 10^3$ \\
	\texttt{mt1e7\_bh1e6} & -- & -- & -- & $2.2$ & $3.7$ & $1.2\times 10^4$ & $8.2\times 10^4$ & $2.4\times 10^3$ \\\vspace{2pt}
	\texttt{mt1e7\_bh5e5} & -- & -- & -- &	$2.1$ & $3.1$ & $1.2\times 10^4$ & $7.4\times 10^4$ & $2.3\times 10^3$ \\
	
	\texttt{mt5e6\_bh2.5e6} & $0.25$ & $9.8$ & $7.9 \times 10^4$ & 	$-$ & $-$ & $-$ & $5.3 \times 10^4$ & $6.1 \times 10^3$\\
	\texttt{mt5e6\_bh1e6} & $0.17$ & $0.9$ & $5.7\times 10^3$ & $1.7$ & $3.5$ & $1.1\times 10^4 $ & $7.6 \times 10^4$ & $3.6 \times 10^3$ \\
	\texttt{mt5e6\_bh5e5} & -- & -- & -- & $1.3$ & $2.7$ & $8.1\times 10^3 $ & $8.3\times 10^4$ & $2.8 \times 10^3$\\
	\texttt{mt5e6\_bh2.5e5} & -- & -- & -- & $1.3$ & $2.9$ & $7.7\times 10^3$ & $8.3\times 10^4$ & $2.3\times10^3$
\end{tabular}\vspace{3pt}
        \end{minipage}%
	{\footnotesize Column~1: run name; column~2: inner radius of the disk in $\pc$; column~3: outer radius of the disk in $\pc$; column~4: disk mass in $\msun$; column~5: inner radius of the ring in $\pc$; column~6: outer radius of the ring in $\pc$; column~7: ring mass in $\msun$; column~8: mass of formed stars in $\msun$; column~9: mass accreted by the \SMBH. All quantities refer to the snapshots presented in Figures~\ref{fig:grid} and \ref{fig:omaggio}.}
        \end{table*}
\makeatletter\onecolumngrid@pop\makeatother
\clearpage

\makeatletter\onecolumngrid@push\makeatother
\begin{figure*}[!hbpt]
  \begin{minipage}{\linewidth}
	\myfloatalign
	\includegraphics[width=0.8\linewidth]{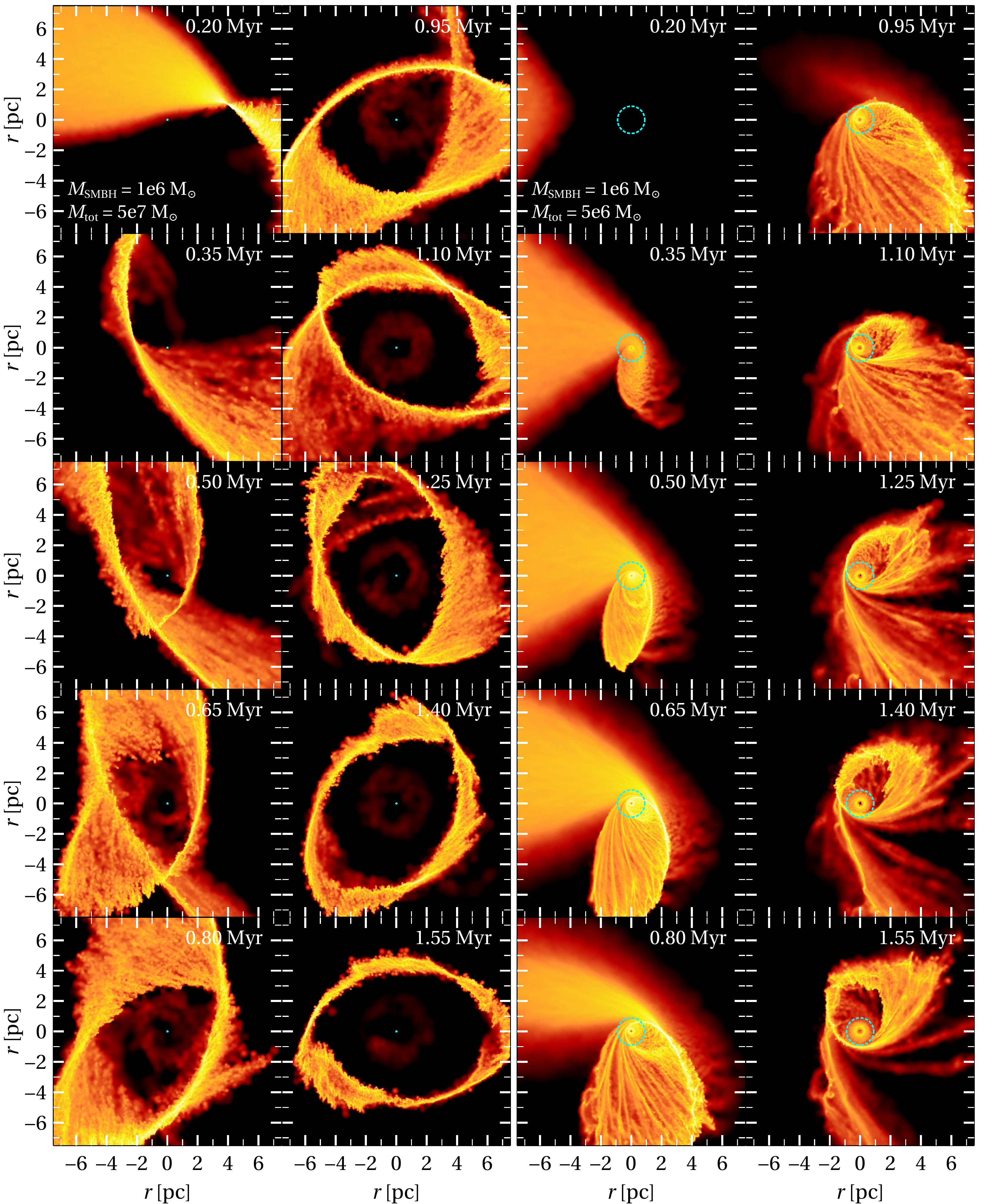}
  \end{minipage}%
	\caption{Color-coded density map of gas in run \texttt{mt5e7\_bh1e6\_hr} (first and second column) and run \texttt{mt5e6\_bh1e6\_hr} (third and fourth column), projected to the $x$-$y$ plane of the simulation. From top to bottom and left to right: $t = 0.4$, $0.55$, $0.7$, $0.85$, $1$, $1.15$, $1.3$, $1.45$, $1.6$ and $1.75 \myr$. The dashed cyan circle indicates the \SMBH radius of influence $R_{\rm SOI}$, which is $0.06$ and $0.35 \pc$ in run \texttt{mt5e7\_bh1e6\_hr} and \texttt{mt5e6\_bh1e6\_hr}, respectively. 
	}
	\label{fig:evol}
\end{figure*}
\makeatletter\onecolumngrid@pop\makeatother

Figure~\ref{fig:evol} shows the time evolution of the gas cloud in run \texttt{mt5e7\_bh1e6\_hr} ($f_{\rm SMBH} = 0.02$, first and second column) 
and run \texttt{mt5e6\_bh1e6\_hr} ($f_{\rm SMBH} = 0.2$, third and fourth column). 
Although the \SMBH mass is the same in both runs, the gas follows a very different evolution. 

In run \texttt{mt5e7\_bh1e6\_hr} the gas cloud is disrupted more rapidly due to the higher total mass ($M_{\rm tot} = 5\times 10^7\msun$). During the first $0.5\myr$ the gas is stretched into a nearly-radial streamer, in which high-density clumps are formed. Afterwards, the gas begins to fall back, intersecting the part of the cloud that is still falling towards the center. 

At $1\myr$ the head of the stream is undergoing a third pericenter passage which is slightly off-set from the first. The streamer follows a rosette-like orbit, which causes it to self-interact and lose eccentricity. At ${\sim}1.7 \myr$ the gas cloud has been completely disrupted and has given birth to a clumpy, eccentric ring. The ring shows a large cavity since little gas is captured by the \SMBH.


\makeatletter\onecolumngrid@push\makeatother
\begin{figure*}[htbp]
  \myfloatalign
    \begin{minipage}{0.47\linewidth}
      \includegraphics[width=\linewidth]{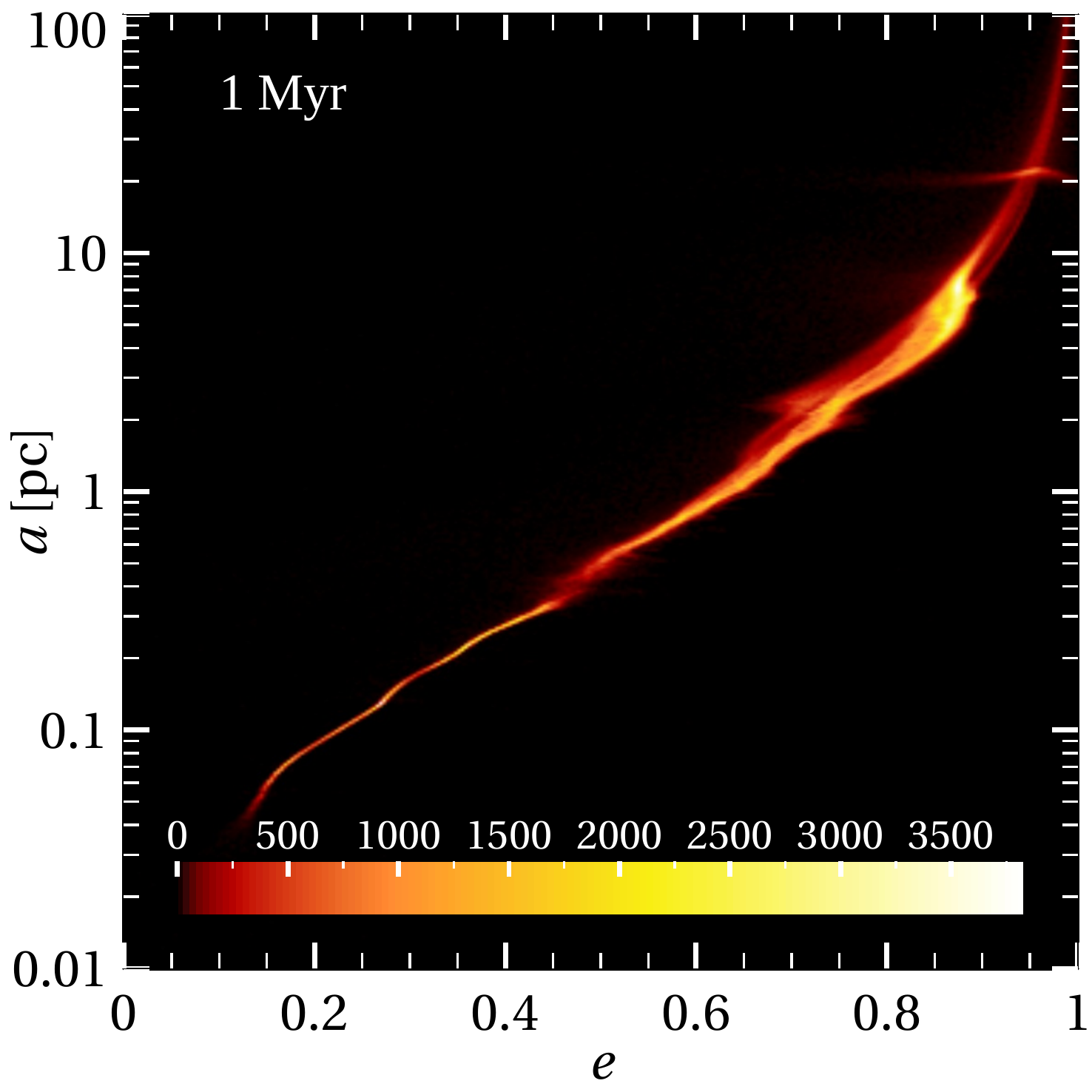}
      \end{minipage}
     \begin{minipage}{0.47\linewidth}
	\includegraphics[width=\linewidth]{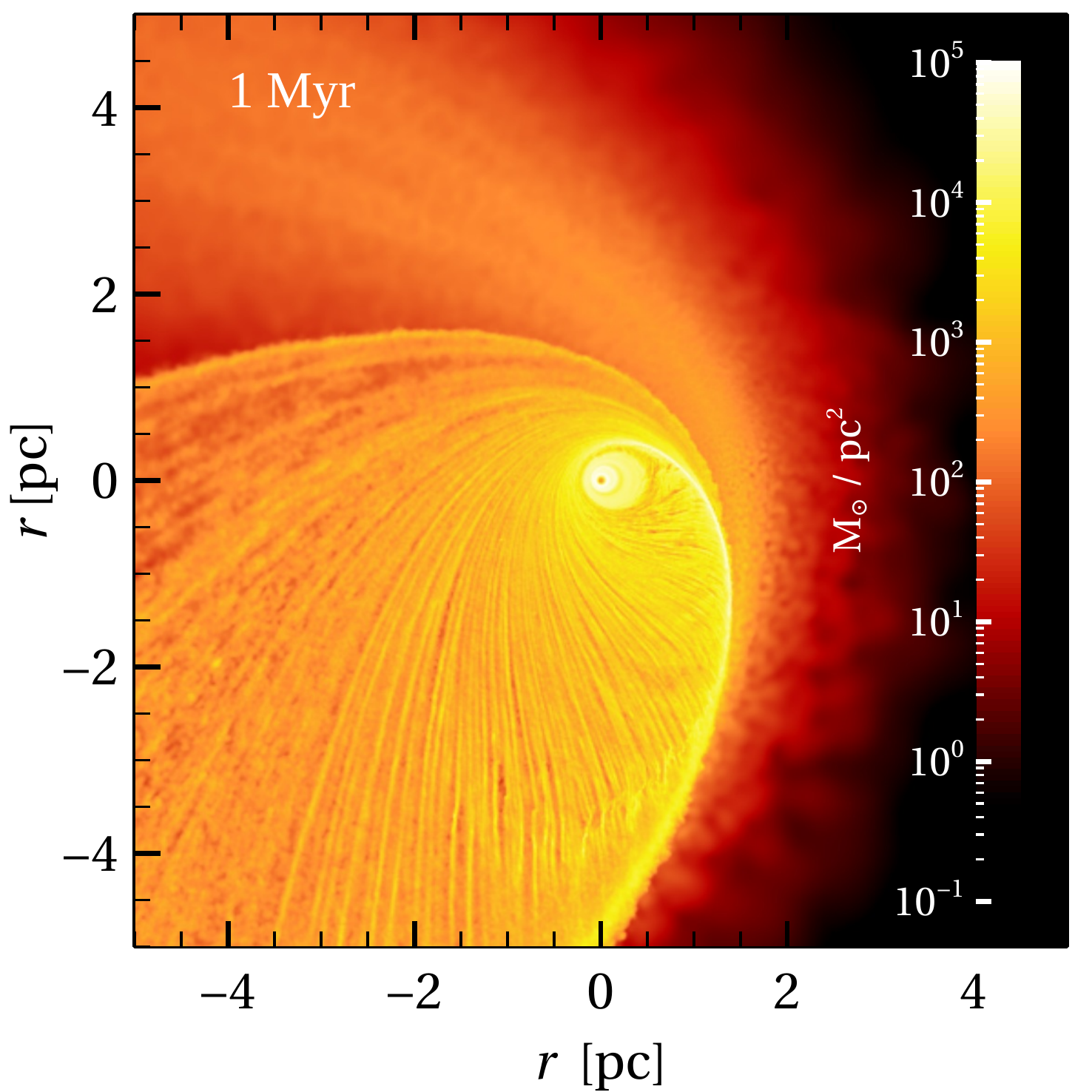}
    \end{minipage}
	\caption{Left-hand panel: semimajor axis versus eccentricity for gas particles for run \texttt{mt1e7\_bh5e6\_hr} ($f_{\rm SMBH} = 0.5$) at $1 \myr$. The color scale indicates the number of gas particles.  Right-hand panel: color-coded, projected density map of gas in the $x$-$y$ plane for the same run at the same snapshot.
	}
	\label{fig:ae}
\end{figure*}

\makeatletter\onecolumngrid@pop\makeatother

Conversely, in run \texttt{mt5e6\_bh1e6\_hr}, the trajectory of the infalling gas is deflected by the \SMBH gravity and the gas winds up around the \SMBH, forming a flattened disk ($0.6\myr$). 
The eccentricity of the disk increases with the distance from the \SMBH. The shear arisen at pericenter leads to a rapid circularization of the inner part of the disk, which decouples from the outer part.

At $1.3 \myr$, there are two distinct structures: a small disk of ${\sim}0.2\pc$ radius around the \SMBH and an outer eccentric ring. 
The external ring is composed of several streamers. The ring precesses because of the cusp potential, so that its pericenter advances along the orbit (in Figure~\ref{fig:evol} it moves clockwise). 
The cloud is completely disrupted at ${\sim}1.7\myr$. The final outcome is a self-interacting, eccentric ring outside $R_{\rm SOI}$ and a dense disk inside $R_{\rm SOI}$.

{

The disk that forms around the \SMBH has increasing eccentricity for increasing semimajor axis, as shown in the left-hand panel of Figure~\ref{fig:ae}. This feature is an effect of shear viscosity that transports angular momentum outwards, causing the inner parts of the disk to circularize faster. 
Neighboring streamlines in the disk intersect because of the eccentricity gradient. This gives rise to shocks in the disk (Figure~\ref{fig:ae}, right-hand panel). The shocks tend to damp the eccentricity of the disk, although full circularization ($e=0$) is never attained in any of the simulations.

About $2$--$7\%$ of the gas is accreted by the \SMBH in all simulations. The accreted mass for each simulation is summarized in Table~\ref{tab:gasres}. The lower $f_{\rm SMBH}$, the less the accreted mass: about $6$--$7\times 10^3 \msun$ are accreted in the simulations with $f_{\rm SMBH} = 0.5$, while $1$--$2\times 10^3 \msun$ are accreted in the simulations with $f_{\rm SMBH} = 0.05$. This value is to be considered as an upper limit since the accretion physics is not resolved at the scale of our simulations.
}

\section{Discussion}\label{sec:discussion}

The simulations show that the evolution of gas changes drastically depending on the relative masses of \SMBH and \NSC. The mass ratio between \SMBH and \NSC determines the radius of the \SMBH sphere of influence, $R_{\rm SOI}$ (see fifth column of Table~\ref{tab:ic}), the region in which the dynamics becomes Keplerian. The simulations indicate that gas inside $R_{\rm SOI}$ exhibits a disk morphology, whereas gas settling further out likely forms an eccentric, clumpy ring with characteristics similar to the \CNR in the \GC.

\begin{figure*}[htbp]
  \myfloatalign
    \begin{minipage}{\linewidth}
      \includegraphics[width=\linewidth]{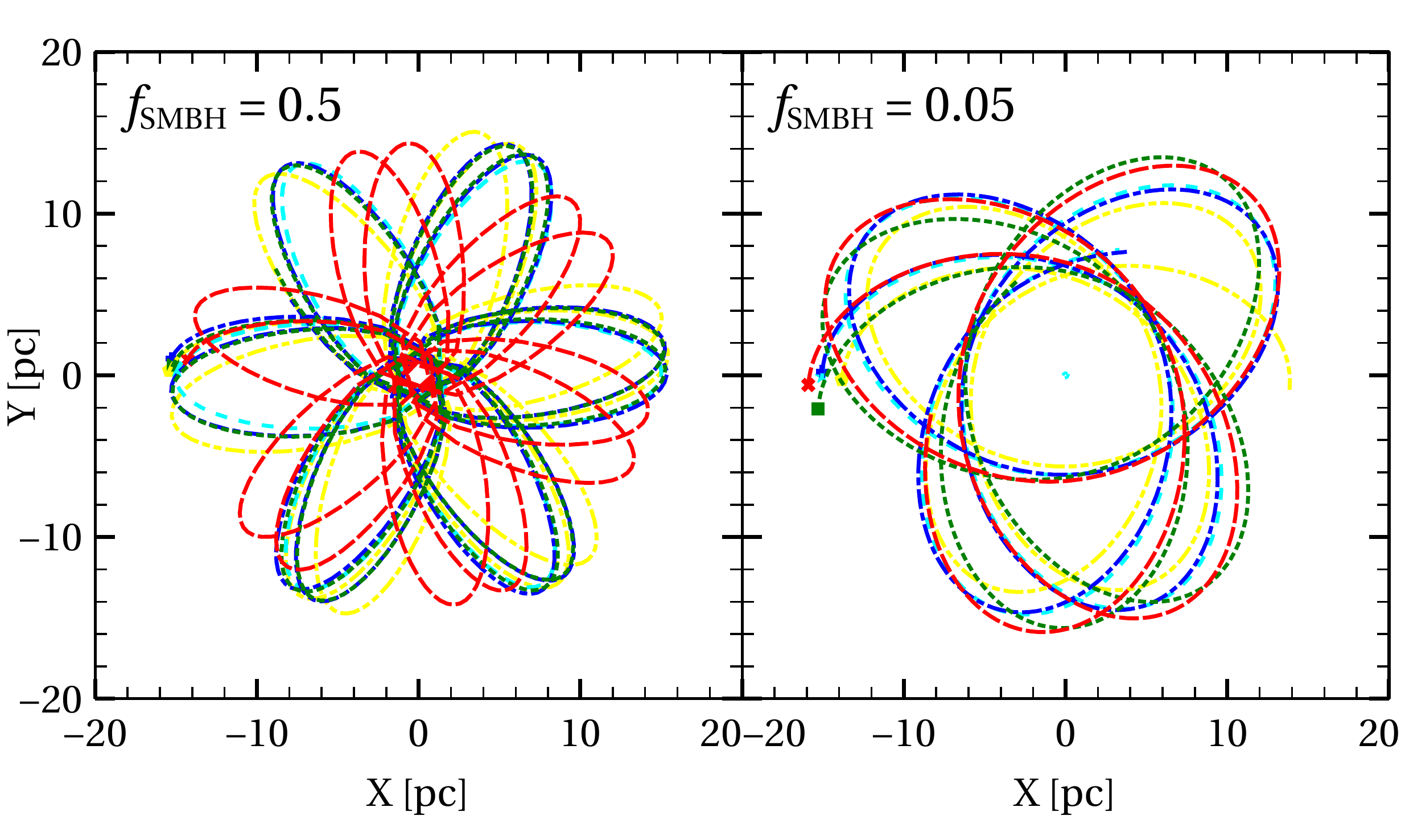}
      \end{minipage}
	\caption{Trajectories of 5 test particles in the gravitational potential of the \SMBH and \NSC. 
		Left-hand panel: $M_{\rm SMBH} = 2.5\times 10^7 \msun$, $M_{\rm tot} = 5\times 10^7 \msun$ ($f_{\rm SMBH} = 0.5$). Right-hand panel: $M_{\rm SMBH} = 2.5\times 10^6 \msun$, $M_{\rm tot} = 5\times 10^7 \msun$ ($f_{\rm SMBH} = 0.05$). Each particle has a different color and line type. The initial position is close to $X,Y = (-15,0) \pc$ and it indicated by a marker. In both panels the particles are evolved for $3 \myr$.
	}
	\label{fig:toy}
\end{figure*}

{
This is due to the way the tidal potential compresses and shapes the cloud as it gets disrupted. Gas captured in the \SMBH sphere of influence tends to wind up around the \SMBH, following eccentric Keplerian orbits. This is apparent in the left-hand panel of Figure~\ref{fig:toy}, which shows the trajectory of neighbouring test particles in the gravitational potential of the \SMBH and the \NSC. If the \SMBH potential is dominant, the test particles move on neighbouring Keplerian orbits, forming a precessing disk.

In this case, the gas undergoes circularization through shocks and shear between intersecting disk streamlines. 
A peculiar feature of this formation mechanism is that the eccentricity of the disk increases from inside out (left-hand panel of Figure~\ref{fig:ae}) and the apsis lines are mutually aligned.


On the other hand, gas that falls into the potential well of the \NSC without reaching $R_{\rm SOI}$ is stretched into a nearly radial streamer. This occurs because the gravitational acceleration exerted by the \NSC scales as $r^{1-\gamma}$, where $\gamma$ is the density power-law index. Since $\gamma$ ranges from $1.2$ to $1.75$, the acceleration induced by the \NSC increases very slowly with $r$ and is not able to strongly alter the orbit of the streamers.
This is clear from the right-hand panel of Figure~\ref{fig:toy}, which shows the trajectory of a test particle in a \NSC-dominated \GN. Test particles follow a wide, self-intersecting rosette orbit. Circularization is induced by the shocks occurring at these self-interactions.
}



This mechanism can naturally explain why the \CNR in the \GC presents an inner cavity and does not extend below $1.5$-$2\pc$ radius \citep{oka11,liu12,liu13,mil13,smi14,har15,tak17,mil17,san17}.

Using the mass profile of \citet{gen03a} for Milky Way's \NSC and the mass estimate of \citep{gil17} for Sgr~A*, the \SMBH influence radius turns out to be $R^{\rm GC}_{\rm SOI}\simeq 0.4 \pc$, much smaller than the inner edge of the \CNR.

Another implication is that the formation of compact, \CNR-like rings is not expected in \GNs lacking a \NSC, since the stellar potential would be too shallow to tidally disrupt a molecular cloud. 
These findings suggest that the radius of a \CNR-like ring in a nucleated \GN could be used as an upper limit of the \SMBH influence radius (and thus of the \SMBH mass), in the hypothesis that the gaseous ring formed according to this mechanism.

Furthermore, $R^{\rm GC}_{\rm SOI}\simeq 0.4 \pc$ is also the outer edge of the so-called clockwise (CW) disk, the disk of young stars around Sgr*A, according to \citeauthor{bar09} (2009, \cite[but see also][who estimate the outer edge at $0.13\pc$]{yel14}). The \CW disk might have originated from the fragmentation of an eccentric disk of gas \citep{nay07,bon08,map08,hob09,ali11,map12,luc13}. This picture is consistent with the scenario presented here, in which an eccentric gaseous disk is expected to form at $r\lesssim R_{\rm SOI}$ if a gas cloud penetrates the \SMBH sphere of influence.

These results also indicate that it is difficult to form a gaseous disk close to the \SMBH without forming a ring outside $R_{\rm SOI}$, unless the cloud is compact enough to fit entirely inside $R_{\rm SOI}$. 
This strengthens the idea that both the \CNR and the \CW disk formed from a single molecular cloud disruption episode \citep{map16a}. 

\makeatletter\onecolumngrid@push\makeatother
\makeatletter\onecolumngrid@pop\makeatother

Interestingly, the CO brightness distribution in the nucleus of two nearby galaxies, NGC3665 and NGC4429, shows a central gap, which roughly coincides with the estimated $R_{\rm SOI}$ of the central \SMBH (see figures~6 of \citealt{oni17} and \citealt{dav18}). 
Based on the results presented here and in \citet{map16a}, a possible explanation for the deficit of gas inside $R_{\rm SOI}$ is the lack of molecular clouds with angular momentum low enough to be captured by the \SMBH gravity and form a disk.

However, the lack of CO emission in the $R_{\rm SOI}$ region of NGC3665 and NGC4429 can be due also to changes of composition in the molecular gas. The simulations presented here do not take into account the evolution of the gas chemistry. In a follow-up study, we will include non-equilibrium chemistry, which will allow us to follow the formation and destruction of H$_2$ and CO self-consistently throughout the simulations. A systematic comparison between our models and observations of other nearby \GNs will be discussed in a forthcoming paper.

\subsection{Caveats}

{
In this study, we did not examine the impact of the initial orbit and intrinsic properties of the molecular cloud on the formation of circumnuclear gas, which was instead investigated in \citet{map16a}. \citet{map16a} found that the initial angular momentum of the cloud determines the radius of the outer rings. A higher initial velocity $v_{\rm i}$ will lead to a larger outer ring, and viceversa. Therefore, we expect that the initial velocity will determine whether gas can reach $R_{\rm SOI}$ and form a disk or not.

Similarly, we expect some dependence on the initial molecular cloud size since the cloud size determines the spread in initial angular momentum of individual gas particles. A larger gas cloud will allow for the formation of a inner disk even for higher total angular momentum, while a more compact cloud will not deposit gas close to the \SMBH unless the initial angular momentum is already very low. More discussion about the initial conditions can be found in Appendix~\ref{sec:app1}. 

}

{
Given enough time, the gas viscosity may affect the radial density profile of the disk inside $R_{\rm SOI}$. We computed the viscous timescale $t_{\rm visc}$ from the properties of the gas in the simulations. The viscous timescale of the inner disk ranges from $4.5 \myr$ to $165\myr$ for a viscosity parameter $\alpha = 0.1$, while for the rings is longer than $200 \myr$. This is much longer than the duration of our simulations.
}

In our simulations, the density profile of the \NSC is modelled as in Milky Way's \NSC, i.e. as broken-power law cusped profile \citep{gen03b}. While this is a conservative choice, the density profile of extra-galactic \NSCs might be different. Present-day observations are not able to resolve in detail the luminosity profile at the very centre of \NSCs, and in recent surveys the surface brightness profiles of \NSCs were fitted with a cored \citet{kin62} model \citep{geo14,geo16}. 

Even so, a \SMBH embedded into a star cluster is expected to develop a stellar cusp in about a relaxation time, which can be below one Hubble time for \NSCs \citep{bah76}. \citet{bah77} predicted the cusp power-law index to be between $1.5$ and $2$ for a realistic multi-mass cluster. This is consistent with the values adopted in this work.

Moreover, since the mechanism described in this work arises from the shape of the potential rather than its overall depth (i.e. total mass of \SMBH plus \NSC), it is expected to hold also for mass regimes not probed by the simulations ($M_{\rm tot}\gtrsim 10^7 \msun $). 

A missing ingredient in this study is the feedback from stars formed in the course of the simulation. The main results presented here are based solely on the dynamics of gas, which could be in principle affected by supernovae, photoionization and outflows from protostars \citep{pel12,dal15}.

While the cusp of the \NSC is mainly composed of old stellar population, young stars can form from the infalling gas cloud.
Indeed, in our simulations the gas fragments and forms stars (or pre-stellar cores), modelled as sink particles. We will study the dynamics and evolution of stars formed in the simulations in a forthcoming paper.
 
Nonetheless, the bulk of star formation occurs at $1\myr$, when the gas has already settled long before the first supernovae may explode. In addition, most stars quickly decouple from their parent gas stream and form a distinct, spatially separated structure. Therefore, their impact on the gas through stellar feedback would be limited. 
More details on the dynamics of stars formed in the simulations will be presented in our next work.

More importantly, the cloud is quickly compressed by tidal forces into streams of dense gas. 
To alter the gas dynamics, the stellar feedback has to induce a velocity comparable to the gas orbital velocity in the \GN potential, which exceeds ${\sim}100\kms$. We thus expect the impact of stellar feedback on the molecular gas to be limited. 

A massive O-type star embedded in the gas of our simulations would have Str\"omgren sphere of radius ${\lesssim}5\times 10^{-3}\pc$.
Although the Str\"omgren radius is rather small compared to the typical size of the gas structures, a fraction of the molecular gas might still get ionized by the stellar radiation. Ionized gas is much more sensible to radiative forces and as such it might display significant deviations from the molecular gas dynamics. One example is the so-called minispiral, a complex of ionized gas filaments that resides in the cavity of the \CNR in the \GC. 

These aspects will be tackled in a forthcoming work, which will include a better treatment of the gas chemistry. In the present paper, we focus only on the dynamics of molecular gas, which is also a better tracer of the underlying gravitational potential.

\section{Summary}\label{sec:conclusions}

We have investigated the formation of circumnuclear disks/rings in \GNs with properties different from those of the \GC, by means of \SPH simulations. We simulated the infall and disruption of a molecular gas cloud towards the central parsecs of a \GN, composed of a \NSC and a \SMBH.

We find that the mass ratio between the \SMBH and the \NSC has a deep impact on the dynamics of circumnuclear gas. Specifically, circumnuclear gas exhibits different morphology depending on whether it settles inside or outside the radius of influence $R_{\rm SOI}$ of the \SMBH.

An extended gaseous disk forms only within $R_{\rm SOI}$, where the gravity of the \SMBH dominates over that of the \NSC.
Gas that falls within $R_{\rm SOI}$ winds up around the \SMBH forming a flattened, eccentric disk. The disk is asymmetric and has an eccentricity that increases with increasing semimajor axis. The disk undergoes circularization due to the eccentricity gradient, which makes neighboring streamlines intersect and shock.

In contrast, compact gaseous rings form only outside the influence radius of the \SMBH.
Gas that falls outside $R_{\rm SOI}$ is stretched into a nearly-radial streamer by the tidal potential of the \NSC. The streamer follows a rosette-like orbit and undergoes circularization through self-interaction. Eventually, a clumpy, eccentric ring (or annulus) at radii larger than $R_{\rm SOI}$.

The different evolution of the gas inside and outside $R_{\rm SOI}$ can explain why the inner edge of the \CNR in the \GC is at large distance from Sgr~A* sphere of influence $R^{\rm GC}_{\rm SOI}\simeq 0.4 \pc$.

These findings indicate that the formation of compact rings of gas naturally occurs in the nuclear regions dominated by the gravity of the \NSC. A remarkable implication is that the inner radius of circumnuclear rings can be used to infer an upper limit to the \SMBH sphere of influence. 



\acknowledgements
We acknowledge the anonymous referee for the constructive comments that helped us improving the manuscript.
We thank Sandro Bressan, Romain Teyssier, Roberto Capuzzo Dolcetta, Monica Colpi and Andi Burkert for interesting discussions. 
This work was supported by JSPS KAKENHI Grant Number 17F17764. AAT and MM acknowledge financial support from INAF through grant PRIN-2014-14 (Star formation and evolution in galactic nuclei). MM and AB acknowledge financial support from Fondation MERAC. The initial conditions were generated using the AMUSE framework \citep{por09,por13,pel13}. All plots were made with the Veusz plotting package.
The simulations were run on the Ulysses cluster at SISSA.

\appendix
\section{Impact of cloud size and orbit}\label{sec:app1}

\begin{figure*}[!htpb]
  \begin{minipage}{\linewidth}
	\myfloatalign
	\includegraphics[width=\linewidth]{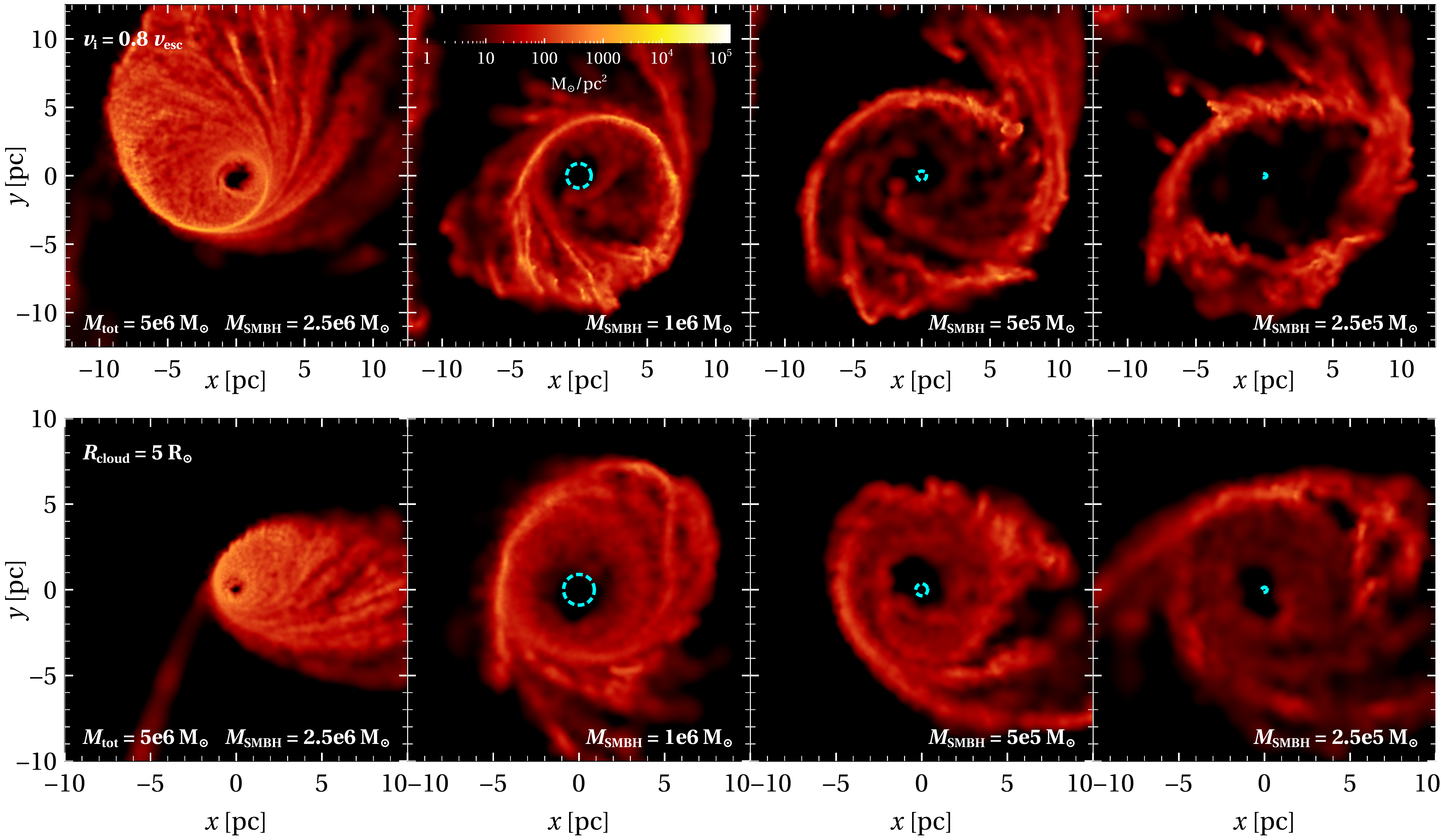}
  \end{minipage}%
	\caption{Color-coded, projected density map of gas in the $x$-$y$ plane for the 8 supplementary simulations (see Table~\ref{tab:icapp}). 
		Top and bottom rows show simulations with $v_{\rm i}=0.8 v_{\rm esc}$ and $R_{\rm cloud} = 5 \pc$, respectively. From left to right, each column has $f_{\rm SMBH} = M_{\rm SMBH} / M_{\rm tot} = 0.5, 0.2, 0.1$ and $0.05$. All simulations consist of $M_{\rm tot} = M_{\rm cusp} + M_{\rm SMBH} =5\times10^6 \msun$. The dashed cyan circle indicates the \SMBH radius of influence $R_{\rm SOI}$. Top (bottom) row corresponds to a time of $3$ ($4.5\myr$) from the start of the simulations. $R_{\rm SOI}$ is not shown in the left-hand panels because it is larger than the box. 
	}
	\label{fig:app}
\end{figure*}

{
We have run two sets of simulations to explore the impact of initial size and orbit of the cloud on the final outcome. 
One set features the disruption of a compact cloud of radius $R_{\rm cloud} = 5 \pc$ and the same mass as the other clouds simulated in this paper, while for the second we simulate a cloud on a wider orbit by setting the initial velocity to $v_{\rm i}=0.8 v_{\rm esc}$. For both sets, we choose $M_{\rm tot} = 5\times10^6 \msun$ and $f_{\rm SMBH} = M_{\rm SMBH} / M_{\rm tot} = 0.5, 0.2, 0.1$, for a total of 8 simulations. The initial conditions are listed in Table~\ref{tab:icapp}.
We expect that little gas will be captured by the \SMBH in the simulations with $f_{\rm SMBH}<0.5$ because of the lack of low-angular momentum gas particles.

Figure~\ref{fig:app} shows the projected density map of gas for all 8 supplementary simulations.
For $v_{\rm i}=0.8 v_{\rm esc}$, the outer ring forms at much larger radius than in the simulations with $v_{\rm i}=0.2 v_{\rm esc}$ (bottom row of Figure~\ref{fig:grid}). It is also apparent that the lower the \SMBH mass, the farther out the ring settles down. This indicates that even if the tidal disruption occurs outside $R_{\rm SOI}$, the final radius of the ring can still be affected by the presence of the \SMBH. In none of the simulations with $f_{\rm SMBH}<0.5$ an inner disk forms inside $R_{\rm SOI}$.

In the simulations with $R_{\rm cloud} = 5 \pc$, the cloud takes longer to be disrupted and settle around the \SMBH due to the higher density and compactness. As in the $v_{\rm i}=0.8 v_{\rm esc}$ cases, no inner disk forms in the simulations with $f_{\rm SMBH}<0.5$. The lack of gas within $R_{\rm SOI}$ is to be attributed to the lack of low-angular momentum particles compared to the  $R_{\rm cloud} = 15 \pc$ simulations.

The disk in the $f_{\rm SMBH}=0.5$ simulation (bottom-left panel of Figure~\ref{fig:app}) is comparable in size to the disk in the $R_{\rm cloud} = 15 \pc$ case. On the other hand, the radius of the rings in the simulations with $f_{\rm SMBH} < 0.5$ is larger than in the $R_{\rm cloud} = 15 \pc$ case. This indicates that for $R_{\rm cloud} = 5 \pc$ and $f_{\rm SMBH}<0.5$, the angular momentum transport is less efficient. This might be due to the narrow spread in impact parameter of individual gas particles, which leads to a narrow distribution of the initial angular velocities. Since here the angular momentum transport arises from shear flow, the viscous torque is proportional to the gradient of the angular velocity, which is shallower for a more compact cloud.

Nonetheless, these simulations display the same qualitative behaviour of those presented in Section~\ref{sec:results}, i.e. a ring forms outside the sphere of influence of the \SMBH, while a disk tends to form inside it.
}

\makeatletter\onecolumngrid@push\makeatother
\makeatletter\onecolumngrid@pop\makeatother


\makeatletter\onecolumngrid@push\makeatother
\begin{table*}
  \caption{Properties of the supplemetary simulations.\label{tab:icapp}}
  \begin{minipage}{\linewidth}
    \centering
  \begin{tabular}{lcrlcccc}
    Run & $M_{\rm tot}$ [$\msun$] & $M_{\rm SMBH}$ [$\msun$] &  $f_{\rm SMBH}$ & $R_{\rm SOI}$ [$\pc$] & $v_{\rm i}/v_{\rm esc}$ & $R_{\rm cloud}$ [$\pc$] \\\hline

	\texttt{mt5e6\_bh2.5e6} & $5\times10^6$ & $2.5\times10^6$ & $0.5$ & $>$$10$ & $0.8$ & $15$\\
	\texttt{mt5e6\_bh1e6} & $5\times10^6$ & $1\times10^6$ & $0.2$ & $0.90$ & $0.8$ & $15$\\
	\texttt{mt5e6\_bh5e5} & $5\times10^6$ & $5\times10^5$ & $0.1$ & $0.35$ & $0.8$ & $15$\\\vspace{2pt}
	\texttt{mt5e6\_bh2.5e5} & $5\times10^6$ & $2.5\times10^5$ & $0.05$ & $0.15$ & $0.8$ & $15$\\

	\texttt{mt5e6\_bh2.5e6} & $5\times10^6$ & $2.5\times10^6$ & $0.5$ & $>$$10$ & $0.2$ & $5$\\
	\texttt{mt5e6\_bh1e6} & $5\times10^6$ & $1\times10^6$ & $0.2$ & $0.90$ & $0.2$ & $5$\\
	\texttt{mt5e6\_bh5e5} & $5\times10^6$ & $5\times10^5$ & $0.1$ & $0.35$ & $0.2$ & $5$\\\vspace{2pt}
	\texttt{mt5e6\_bh2.5e5} & $5\times10^6$ & $2.5\times10^5$ & $0.05$ & $0.15$ & $0.2$ & $5$\\
        \end{tabular}\vspace{3pt}
\end{minipage}%
	{\footnotesize 
		Column~1: run name; column~2: mass enclosed in a $10\pc$ radius $M_{\rm tot}$ in $\msun$, composed of the NSC and the \SMBH; column~3: mass of the \SMBH $M_{\rm SMBH}$ in $\msun$; column~4: mass of the \SMBH with respect to total mass enclosed $f_{\rm SMBH}$; column~5: radius of sphere of influence $R_{\rm SOI}$ of the \SMBH in $\pc$; $m_{\rm res}$ is the mass resolution of the simulation in $\msun$.}
\end{table*}
\makeatletter\onecolumngrid@pop\makeatother

  
\bibliography{ms.bib}

\end{document}